\documentclass[twocolumn]{aastex62}
\usepackage{amsmath, amsfonts}
\usepackage{amssymb}
\usepackage{graphicx}
\usepackage{dcolumn}
\usepackage{bm}

\usepackage{color}

\def\bgp{\begin{pmatrix}}
\def\edp{\end{pmatrix}}
\def\bgs{\begin{subequations}}
\def\eds{\end{subequations}}
\newcommand{\order}[1]{\mathcal{O}({#1})}
\def\di{{\mathrm{d}}}

\def\to{\rightarrow}

\def\al{\alpha}
\def\ga{\gamma}

\def\lam{\lambda}
\def\rh{\rho}

\newcommand{\wh}[1]{\mkern 2mu \widehat{\mkern-2mu#1\mkern-2mu}\mkern 2mu}

\newcommand{\be}{\begin{equation}} 
\newcommand{\ee}{\end{equation}}  
\newcommand{\bea}{\begin{eqnarray}}  
\newcommand{\eea}{\end{eqnarray}}

\begin{document}

\title{Hidden-Sector Spectroscopy with Gravitational Waves from Binary
  Neutron Stars}

\author{Djuna Croon}
\affiliation{Department of Physics and Astronomy, Dartmouth College, Hanover, NH 03755 USA}

\author{Ann E.~Nelson}
\affiliation{Department of Physics, Box 1560, University of Washington, Seattle, WA 98195-1560 USA}

\author{Chen Sun}
\affiliation{CAS Key Laboratory of Theoretical Physics, Institute of Theoretical Physics, Chinese Academy of Sciences, Beijing 100190, P.~R.~China}
\affiliation{Department of Physics and Astronomy, Dartmouth College, Hanover, NH 03755 USA}

\author{Devin G.~E.~Walker}
\affiliation{Department of Physics and Astronomy, Dartmouth College, Hanover, NH 03755 USA}

\author{Zhong-Zhi Xianyu}
\affiliation{Department of Physics, Harvard University, 17 Oxford St., Cambridge, MA 02138, USA}
\affiliation{Center of Mathematical Sciences and Applications, Harvard University, 20 Garden St., Cambridge, MA 02138, USA}


\begin{abstract}
We show that neutron star binaries can be ideal laboratories to probe hidden sectors with a long range force. In particular, it is possible for gravitational wave detectors such as LIGO and Virgo to resolve the correction of waveforms from ultralight dark gauge bosons coupled to neutron stars. We observe that the interaction of the hidden sector affects both the gravitational wave frequency and amplitude in a way that cannot be fitted by pure gravity.
\end{abstract}	


\section{Introduction}
\label{sec:introduction}

The LIGO and Virgo collaborations have
observed gravitational waves (GWs) GW170817 from an inspiraling binary
of neutron stars (NSs)~\cite{TheLIGOScientific:2017qsa}.  This signal,
and the associated electromagnetic counterpart
GRB170817a~\cite{Monitor:2017mdv,GBM:2017lvd}
, provide a wealth of information about neutron stars. In particular, the finely measured waveform of GWs with relatively long duration ($\sim$~20~min) could reveal many otherwise hidden information about the physical properties of NSs themselves and of their ambient environment~\cite{Randall:2017jop}, many of which could be originated from a new physics sector \cite{Ellis:2017jgp}. 

In this Letter, we show that such binary NSs can be an ideal
laboratory for probing a long range force mediated by a new ultralight
particle with   inverse mass $m^{-1}$ comparable to or larger than the
binary separation $r$ in natural units. The LIGO band for binary
separation $r$ is roughly $\order{10-1000}$ km, and this translates to
$\order{10^{-13}\sim 10^{-11}}$ eV for mass. A prototypical example of
this long range force is a hidden sector with asymmetric dark matter
(aDM) along with an ultralight mediator, e.g.~a dark photon. The
charged aDM particles can be trapped by a NS during its lifetime or
could have been present in the star at birth. Binary NSs containing
such aDM then feel  a long range force mediated by the massive but
ultralight dark photon.  

Independent of the  detailed mechanism of aDM-trapping in NSs, we shall show, in a fairly model-independent manner, that clean and detectable GW signals can be generated by aDM-carrying neutron star (NS) pairs. In the current work, we only make two mild assumptions about the hidden-sector companion of NSs: 1) For the mechanism to work, it is assumed that the DM comprises a small mass fraction of the NSs. This implies that the attractive gravitational force is stronger than the hidden repulsion on relevant macroscopic scales. 2) The mass profile of DM in NSs does not differ significantly from the neutron star matter, as is the case for a hidden sector without a repulsive interaction~\cite{Rezaei:2016zje}. This ensures that the point mass approximation for binary NSs holds, and that clean and well-defined GW signals can be derived.

The chirping GWs generated by a pair of purely self-gravitating point masses $m_1$ and $m_2$ has the following time-dependent frequency $f_\mathrm{GW}$, to leading order in the post-Newtonian expansion,
\be
\label{fGW}
  f_\mathrm{GW}(t)=\frac{1}{\pi}\bigg(\frac{G \,m_c}{c^3}\bigg)^{-5/8}\bigg(\frac{5}{256}\,\frac{1}{t_0-t}\bigg)^{3/8},
\ee
where $t_0$ is the time of the coalescence and $m_c=(m_1
m_2)^{3/5}/(m_1+m_2)^{1/5}$ is the chirp mass. For GW170817, the chirp
has been measured with rather high precision as $m_c =
1.188\substack{+0.004 \\ - 0.002}
\,M_\odot$~\cite{TheLIGOScientific:2017qsa}. 

The chirp  signal is subject to corrections from hidden sectors, of which we identify two important effects with the above mild phenomenological assumptions. The first effect is due to the Yukawa potential between the two NSs coming from the exchange of a massive dark photon. In the case that dark photon mass $m_V$ lies within the LIGO band, the Yukawa repulsion between to two NSs is virtually absent when the binary separation $r\gg m_V^{-1}$, but behaves like Coulomb repulsion when $r\ll m_V^{-1}$. The Coulomb repulsion would affect the observed chirp mass of the binary, and therefore the Yukawa potential will generate a characteristic shift in chirp mass during the inspiraling phase that can in principle be observed by LIGO and Virgo detectors. 

The second effect is from the fact that such inspiraling NS binaries generate dark radiation, so long as the wavelength of the radiation is much greater than $m_V^{-1}$. If the two NSs do not share exactly the same (dark) charge-mass ratio, then the dark radiation develops an electric dipole component which is qualitatively different from GWs because the latter start only from quadrupole level, and therefore the dark dipole radiation will generate a distinct correction to GW signals. 


Conversely, in detected events without observed hidden-sector corrections such as in GW170817, bounds can be placed on the relative strength of the long range force between the two NSs, as will be detailed below. Importantly, these constraints hold independently of both the strength and form of the portal coupling, and the DM relic density.  

In the rest of this Letter we first describe a generic model of asymmetric dark matter with an ultralight mediator. 
We then describe in detail the two effects, the dark repulsion and the dark radiation. We summarize with a discussion on the applicability of our results.

\section{A Generic Model}
\label{sec:generic-model}

 Though our analysis of hidden sector corrections to GWs is rather generic and model independent, it is very helpful to illustrate the general point with a simple model of DM charged under $U(1)'$,
\be\mathcal{L} = \mathcal{L}_{V}  +  \mathcal{L}_{\chi} +\mathcal{L}_{mix},\ee
where  $\mathcal{L}_V$ is the vector potential of the gauge fields, and
$\mathcal{L}_\chi$ the Lagrangian of DM,
\bea 
\mathcal{L}_{V} &=&  - \frac{1}{4} V_{\mu\nu}V^{\mu\nu} + \frac{1}{2} m_V^2 V_\mu V^\mu  \\
\mathcal{L}_{\chi} &=& \bar{\chi} \left( \gamma^\mu (i \partial_\mu - g' V_\mu) - m_\chi  \right) \chi  \, .
\eea
Here $m_V$ is the dark photon Stuckelberg mass term, $m_\chi$ is the dark matter mass, $g'$ is the gauge coupling. We will exchange the latter for $\al'=g'^2/4\pi$ throughout the letter.  We make no assumptions on the form of $\mathcal{L}_{mix}$, though traditionally that part of the lagrangian holds the photon-dark photon kinetic mixing terms.  Throughout, we require the dark photon mass to reside within or be lighter than the LIGO-Virgo detection window, $m_V^{-1}> \order{10}$ km.

\section{Dark Repulsion}
\label{sec:dark-repulsion}

Due to the presence of aDM, the two NSs carry like $U(1)'$ charges, such that a repulsive force is generated. The range of the force is determined by the
mass of the dark photon. When the distance of the two NSs is large the interaction can be neglected; when they are within the
effective range $m_{V}^{-1}$, the repulsive dark force is switched
on, and effectively behaves like a Coulomb force at small distance. This is the reason NS mergers are particularly good for
constraining dark photons in the range $m_{V}^{-1} \sim \mathcal{O}(10 - 1000)$ km, with NS
radius being $\mathcal{O}(10)$ km, and separation when the signal enters
LIGO-Virgo window up to $\mathcal{O}(1000)$ km. 

For a binary system of NSs with mass $m_{1,2}$ and dark $U(1)'$ charge $q_{1,2}$, the orbital frequency of the inspiraling binary is given by
\be
\label{omega2}
  \omega^2=\frac{Gm}{r^3}\big(1-\tilde\al'e^{-m_V r}\big),
\ee
where $m=m_1+m_2$ is the total mass of the
binary, $r$ is the binary separation, and $\tilde\al'$ is defined to be,
\begin{align}
\label{alprime}
\tilde\al' \equiv \frac{\al' q_1 q_2}{G m_1 m_2}
=
\frac{ \alpha'}{G} \left (\frac{f}{m_\chi} \right)^2, 
\end{align}
 where $f$ is the fraction of aDM in each NS.
The total energy of the system is,
\be \label{RepEn}
  E=E_G + E_V=-\frac{Gm\mu}{2r }\big(1-\tilde\al' e^{-m_Vr}\big),
\ee
where $\mu=m_1m_2/m$ is the reduced mass. 
%

\begin{figure}[t]
\centering
\includegraphics[width=0.48\textwidth]{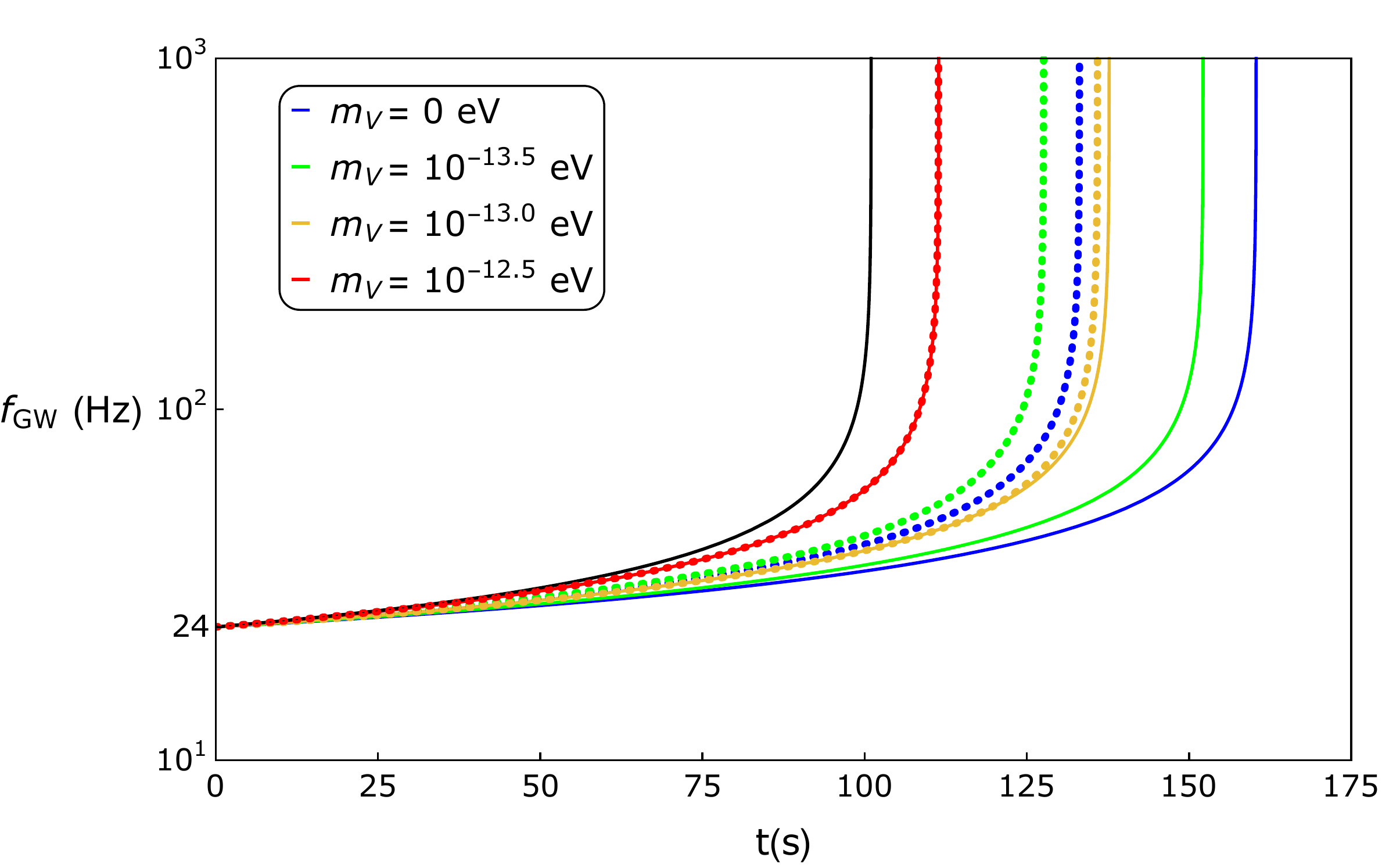}
\caption{
\label{fig:freq-time-diff-mV}The change of frequency-time
  dependence due to the dark boson. Here, the dimensionless coupling in (\ref{alprime}) is set to $\tilde \alpha' = 0.5$ for illustration. 
({ See
  main text for a realistic choice of $\tilde \alpha'$ and the
 description of the signal in the change of shape.})
The NS masses are chosen to be $m_1=1.51\;M_\odot$ and $m_2=1.24\;M_\odot$ respectively, such that the chirp mass before aDM correction corresponds to that of GW170817. The corresponding limits $m_V \rightarrow \infty$ and $\tilde \alpha' \rightarrow 0$ are degenerate and given by the black curve. 
The solid lines correspond to zero charge-mass difference $\gamma$ and thus show the effect of dark repulsion.
The dotted lines correspond to a charge-mass difference of $\gamma =
q_1$, and therefore show the competing effects of dark dipole
radiation and repulsion. It is seen that the correction due to dark
dipole radiation is important for relatively light dark mediators, and
is switched on for $m_V c /\omega<1$ as detailed in the text. }
\end{figure}

The result of the new force gives a distortion of the waveform due to
the fact that the $r$ dependence of Yukawa potential is different from
gravity. To see the point qualitatively, we first note that the Yukawa repulsion is absent when $r>m_V^{-1}$, which corresponds to the early stage of inspiraling in the LIGO band. In this regime the waveform is identical to a purely self-gravitating pair as shown in (\ref{fGW}). On the other hand, when the binary separation reduces below $m_{V}^{-1}$, which corresponds to a later stage in LIGO band, the dark repulsion can be approximated by a Coulomb force with $r^{-2}$ law. { This force is identical with gravity in $r$ dependence and thus the waveform will still be (\ref{fGW}) but with a modified chirp mass due to the Coulumb repulsion. In summary, the apparent chirp mass is,
\be
\label{whmc}
\wh m_c=\left\{\begin{split}&m_c &&(r>m_V^{-1}),\\
&(1-\tilde\al')^{2/5}m_c &&(r<m_V^{-1}).
\end{split}\right.
\ee
The factor $(1-\tilde\al')^{2/5}$ in the second line is found by solving the equation of energy conservation $\dot E=-P_{GW}$, where $E$ is given in (\ref{RepEn}) with $m_V\to 0$ and $P_{GW}$ is the power of quadrupole radiation of GWs and is given by (see, e.g., \cite{Maggiore} for pedagogical introductions and \cite{Randall:2017jop} for a short review),
\be
\label{PGW}
  P_{GW} =\frac{32G\mu^2\omega^6r^4}{5c^5}.
\ee
}
Therefore, we see that the net effect of Yukawa repulsion is to generate a characteristic and observable shift in the chirp mass, if the dark photon mass lies within the LIGO band. { We stress that the shift of the chirp mass is \emph{not} a uniform or time-independent correction, but occurs within a short period of time in the LIGO band when $r$ goes across the Yukawa threshold $m_V^{-1}$ and thus is resolvable by a precise measurement of the chirp mass.

The full waveform corrected by Yukawa repulsion can be worked out again by solving the energy-conservation equation $\dot E=-P_{GW}$ but with the full expression of $E$ in (\ref{RepEn}).} Here it is more convenient to work out $r(t)$ instead of $\omega(t)$. This can be solved analytically to first order in $\tilde{\alpha}'$,
\be
  \frac{\di r}{\di t}\simeq -\frac{64G^3m^2\mu}{5c^5r^3}\bigg[1+\tilde{\alpha}' e^{-m_V r}(m_V r-2)\bigg].
\ee
From this equation we can solve for $r(t)$ to first order in
$\al'$. Let $r(0)=r_0$ be the binary separation when the signal enters
the LIGO-Virgo band then $r(t)$ can be found
\begin{align}
  &t =\frac{5c^5}{64G^3m^2\mu}\bigg[\frac{1}{4}(r_0^4-r^4)+\tilde{\alpha}'\Big(f(r_0)-f(r)\Big)\bigg],\\
  &f(r)\equiv\frac{e^{-m_V r}}{m_V^4}\Big[12+m_V r(2+m_V r)(6+m_V^2r^2)\Big].
\end{align}
With $r(t)$ known, the corrected time dependence of the orbital frequency $\omega(t)$ can be found from (\ref{omega2}), and the corrected chirp signal has frequency $f_{GW}$ and amplitude $A_{GW}$ \cite{Maggiore},
\begin{align} 
\label{fGWt}
f_{GW}(t) = &~\frac{\omega(t)}{\pi}, \\
\label{AGWt}
  A_{GW}(t)=&~\frac{1}{d_L}\frac{2\,G}{c^4}\cdot 2\,\mu\,\omega^2(t)r^2(t),
\end{align}
where $d_L$ is the luminosity distance of the source. The effect on
frequency $f_{GW}(t)$ and amplitude $A_{GW}(t)$ of the GW signal are
shown as solid curves in Figure~\ref{fig:freq-time-diff-mV} and
Figure~\ref{fig:amp-time-diff-mV} respectively. The initial frequency
is set to be consistent with Ref.~\cite{TheLIGOScientific:2017qsa},
which corresponds to different separations of the two stars. { Although the most significant effect of dark repulsion in Figure~\ref{fig:freq-time-diff-mV} is the lengthened signal duration, we stress again that it is the less visible effect of the shifted chirp mass that lead to a distinct signal non-degenerate with the pure gravity case (black curve). Furthermore, we emphasize that the value of $\tilde\al'$ taken in Figure~\ref{fig:freq-time-diff-mV} and
Figure~\ref{fig:amp-time-diff-mV} and much exaggerated only for illustrative purpose. In reality, the value of $\tilde\al'$ is much smaller and we shall show later that the LIGO-Virgo measurement of GW170817 can actually be sensitive to a smaller $\tilde\al'$ down to $\order{10^{-2}}$. 
}

\section{Dark radiation}
\label{sec:dark-radiation}

A separate effect of the hidden sector comes from the fact that a pair of dark 
charged inspiraling NSs radiate dark photons. If the charge-mass ratios of the two stars are not identical,  a net (dark electric) dipole moment stimulates dipole radiation. This dark dipole radiation drains additional energy away from the system and thus affects the waveform of the chirp signal. Importantly, the dipole radiation is stimulated only when the frequency is higher than the dark photon mass, i.e. $m_V c/\omega<1$, and is otherwise quenched. Therefore, the dark photon mass should be $m^{-1}> \order{10^2}$km for the dipole radiation to be generated within the LIGO band. This includes a very light dark photon with $m_V^{-1}>\order{1000\text{km}}$ where the dark photon becomes effectively massless in the Yukawa potential for the whole range of LIGO band, and the
dipole radiation becomes the most important correction. In this case the dark repulsion is well approximated by a Coulomb potential, such that the  repulsive effect is degenerate with the pure gravity case with modified chirp mass $\wh m_c=(1-\tilde{\alpha}')^{2/5}m_c$ and therefore cannot be observed directly.

\begin{figure}[t]
\centering
\includegraphics[width=0.48\textwidth]{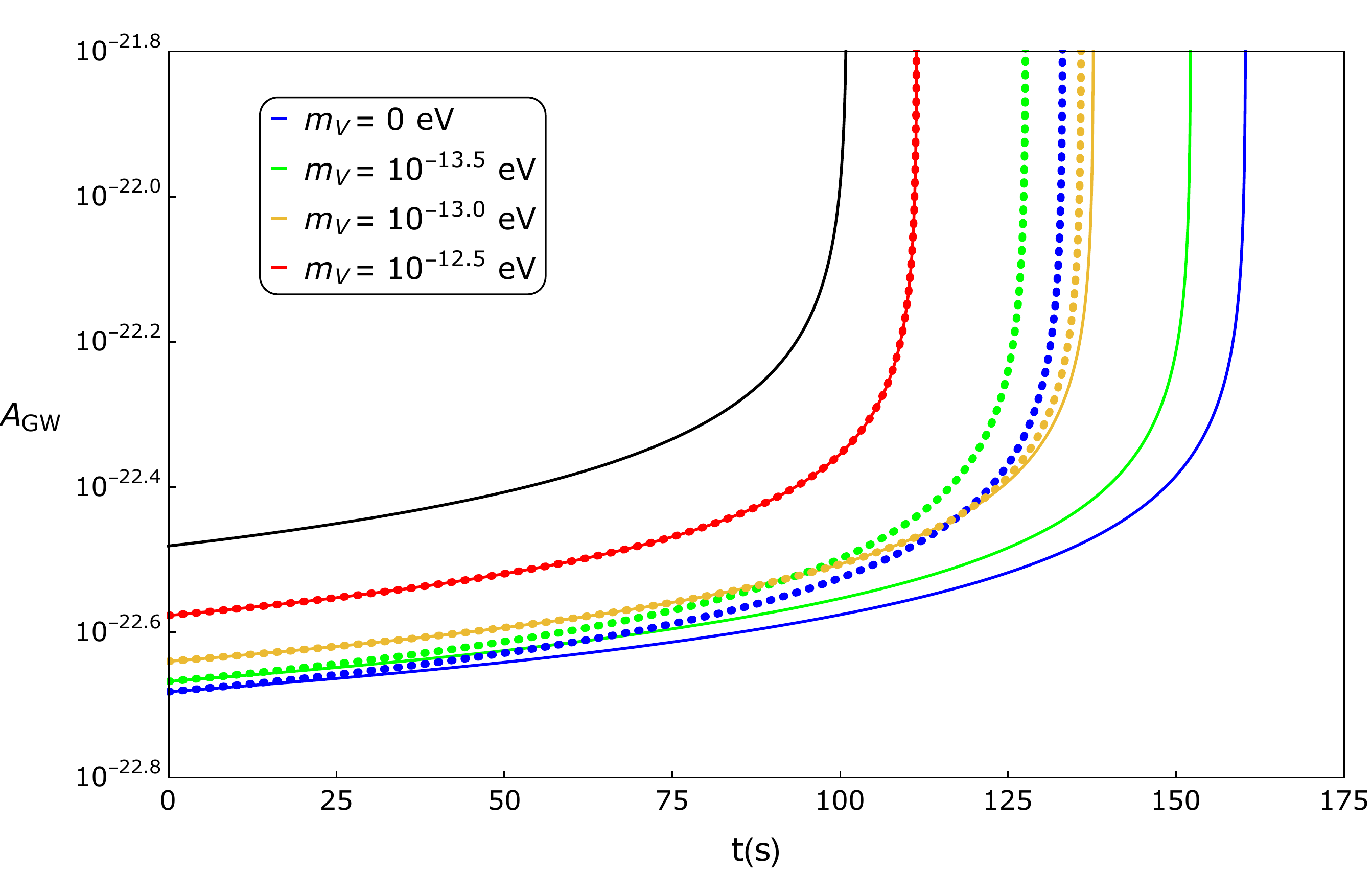}

\caption{
\label{fig:amp-time-diff-mV}The change of Amplitude-time
  dependence due to the dark boson, under the same assumptions as in
  Figure~\ref{fig:freq-time-diff-mV}. The results are normalized such
  that $f_{GW}(0) = 24$ Hz.}
\end{figure}

The total power of dark dipole radiation is given by~\cite{Krause:1994ar},
\be
  P_\mathrm{dark} =\left\{
  \begin{split}&\frac{\al'\ga^2\omega^4r^2}{3c^3}\sqrt{1-\Big(\frac{m_V c}{\omega}\Big)^2} \\ & \times \bigg[1+\frac{1}{2}\Big(\frac{m_V c}{\omega}\Big)^2\bigg] &&(\omega>m_V c),\\ 
  &0 &&(\omega\leq m_V c).
  \end{split}
  \right.
\ee
where $\ga\equiv \mu|\rh_1-\rh_2|$ and where $\rh_i=q_i/m_i~(i=1,2)$
is the charge-mass ratio of each NS. From this it is seen that the effect is absent if both NS carry the same DM mass fraction, as expected. 
The chirp signal is now found from $\dot E=-P_{GW}-P_\mathrm{dark}$, or more explicitly,
\begin{equation}
 \frac{\di\omega}{\di t}=X\,\omega^{11/3}+Y\,\omega^3,
\end{equation}
where 
\begin{align}
X=\frac{96}{5}\bigg(\frac{G\,\wh m_c}{c^3}\bigg)^{5/3},  &&Y=\frac{\al'\ga^2}{c^3\mu}.
\end{align}
The equation can be readily integrated to get the chirp signal $\omega(t)$ and it is clear that the resulted waveform deviates from the standard form in equation~(\ref{fGW}). 
In particular, if we assume the correction is small (which must be for detected events), then we are allowed to expand the result in terms of small $Y$ to get the following modified chirp signal,
\be
  \omega=\bigg(\frac{3}{8X(t_c-t)}\bigg)^{3/8}-\frac{Y}{10X}\bigg(\frac{3}{8X(t_c-t)}\bigg)^{1/8},
\ee
where $t_c$ is the time of coalescence. With $\omega(t)$ and $r(t)$
known, the frequency and amplitude of the signal can again be obtained
from (\ref{fGWt}) and (\ref{AGWt}), which is
shown as dotted curves in Figure~\ref{fig:freq-time-diff-mV} and
Figure~\ref{fig:amp-time-diff-mV}.

When the leading electric dipole is suppressed by the small difference in the charge-mass-ratio, the higher multipoles start to dominate, including the magnetic dipole and electric quadrupole. Again, they are stimulated only if $m_Vc/\omega<1$. In this case, the total power of the magnetic dipole radiation is,
\bea \notag
  P_{M1}&=&\frac{\al'}{12c^5}\frac{\mu^2}{m^2}(m_2\rh_1+m_1\rh_2)^2\omega^6r^4 \\
  &\times& \bigg[1-\Big(\frac{m_V c}{\omega}\Big)^2\bigg]^{3/2}.
\eea
Whereas the total power of the electric quadrupole radiation is,
\bea \notag
  P_{E2}&=&\frac{\al'}{72c^5}\frac{\mu^2}{m^2}(m_2\rh_1+m_1\rh_2)^2\omega^6r^4 \\ &\times& \bigg[1-\Big(\frac{m_V c}{\omega}\Big)^2\bigg]^{3/2} \bigg[1+\frac{22}{15}\Big(\frac{m_V c}{\omega}\Big)^2\bigg]^{3/2}.\,\,\,
\eea
Both are zero when $\omega\leq m_V c$. Not surprisingly, they are
degenerate with the leading order GW radiation and thus do not
generate independent corrections to the original chirp signal. 

\section{Comparison with Scalar Tensor Theory}
\label{sec:comp-with-scal}

It is known that a large class of gravitational interactions that deviate
from the general relativity prediction can be described in terms of the
scalar-tensor theory. As one type
of scalar-tensor theory, Brans-Dicke theory has constant scalar
charge. Therefore, it receives universal bounds {and can potentially
affect the NS's equation of state \cite{Will:1989sk}. The current
bounds on Brans-Dicke theory come from solar system tests and lack of
dipole radiation from certain binary pulsar systems.} It is
believed that within in current bound Brans-Dicke theory is
not likely to generate a dipole radiation large enough to be observed
in GW experiments \cite{Sampson:2014qqa}.
However, there is another class of scalar-tensor theories in
which the scalar charge can be dynamically acquired
\cite{Damour:1992we,Damour:1993hw,Barausse:2012da,Damour:1996ke,Palenzuela:2013hsa}.
In particular, Ref.~\cite{Barausse:2012da} shows a way that the scalar
charge can be dynamically generated in the binary merger system after
the orbital binding energy reaches a threshold. Rigorous
bounds on the theory parameter $\beta_{ST}$  from NS mergers are
demonstrated in Ref.~\cite{Sampson:2014qqa} for massless scalar case
and Ref.~\cite{Sagunski:2017nzb} for massive scalar.
In particular, the
scalar charge defined in Ref.~\cite{Sagunski:2017nzb}  can be mapped to our scalar charge as
$\tilde \alpha' = -\alpha^2 = -2 \beta^2$. For example, in $f(R)$
gravity, $\beta=6^{-1/2}$ corresponds to $\tilde \alpha' =
-1/3$. While \cite{Sagunski:2017nzb} considers the scalar interaction
being attractive, we consider the repulsive case due to NS binary
carrying like dark charge, \textit{i.e.} $\tilde \alpha' > 0$. 
In addition, even though Ref.~\cite{Sampson:2014qqa} studies a massless
scalar, the acquired scalar charge in the inspiral phase effectively
switches on the interaction, which is similar to the effect in our
model. We note that, while in the scalar tensor theory one can have both
dipole radiation and `turn-on' signal in the inspiral phase due to
scalarization, which can be captured with ppE$_\theta$ template
\cite{Sampson:2014qqa,Sampson:2013lpa}, our model only permits either
of the two signals at a time, depending on the mass of the mediator. 
{ We also note in passing that, another difference between the
scalar tensor theory and our model lies in the fact that the dark
force in our model would not affect the black hole mergers.}

\section{Discussions}
\label{sec:discussions}

We can distinguish two different observational windows for our binary
neutron star ``spectrometer"  depending on the mediator mass.
Firstly, in the range of  $\order{10}$~\text{km}~$\lesssim (m_V)^{-1}
\lesssim \order{1000}$ km, for which dark radiation is mostly
suppressed. As can be seen from \eqref{RepEn}, the dark repulsion is
suppressed in the early stages of the merger, but behaves effectively
as Coulomb repulsion at late stages. This growth translates into
different apparent values of chirp mass in early and late stages,
$m_c$ and $\wh m_c$, which differ by a factor of
$(1-\tilde{\alpha}')^{2/5}$.  
Therefore, for models with large enough $\tilde{\alpha}'$, it will
become impossible to fit the whole waveform with a single standard
template with a unique chirp mass. It will be necessary to use two
templates with different masses $m_E$ and $m_L$ to fit the early
waveform and late waveform, respectively. The difference $m_E-m_L$
then gives the value of $\tilde{\alpha}'$, and the place where both
templates fail to fit gives the mass of the dark photon.  For a
rigorous waveform analysis and comparison of different templates, we  refer to
\cite{Sampson:2014qqa,Sennett:2016rwa,Sagunski:2017nzb,Shao:2017gwu,Sennett:2017lcx,Yunes:2016jcc}. We leave a detailed analysis for future work.

 Without going into a full waveform analysis, we outline our 
  estimate on the size and detectability of $\tilde \alpha'$. From (\ref{whmc}) it is straightforward to see that the bound
is given by $\tilde{\alpha}' \lesssim \frac{5}{2}(m_{c,i}-m_{c,f})/m_c$,
 where $ m_{c,i}\; (m_{c,f})$ is the initial (final) chirp mass, respectively. A
large enough $\tilde \alpha'$ corresponds to a large change of $m_c$
in the signal. 
We note that a rigorous bound can be
  drawn only if one analyzes the data using a template with $\tilde
  \alpha'$ parameter. 
A Markov chain
  Monte Carlo (MCMC) simulation \cite{Sagunski:2017nzb} using 
  \texttt{emcee} \cite{ForemanMackey:2012ig} shows that
  $|\tilde{\alpha}'|\sim 1/3$ is detectable at LIGO.
This is to be 
compared with the ``trivial'' stability bound for a charged NS star,
which is given by setting $E=0$ in (\ref{RepEn}) and roughly
translates to $\tilde{\alpha}'<1$~\cite{Ray:2006qq,Ray:2003gt}. 
  Again, we note that this bound is valid only if one assumes a mediator in
  the range of $\order{10}$~\text{km}~$\lesssim (m_V)^{-1}
\lesssim \order{1000}$ km. Failure to observe a change in the chirp mass
can be interpreted as either too small an $\tilde \alpha'$ or $m_V$
beyond the neutron star ``spectrometer" window.  We
 also  note that the quantity $\tilde\al'$ is the product of the coupling
  strength $\al'$ and the charges carried by the binary $q_1q_2$, and
  it is in principle possible that $\tilde\al'$ is different for
  different binaries due to different $q_1q_2$. Therefore, the
  $\tilde\al'$ is to be measured event-wise, and thus the improvement
  of upper limit on $\tilde\al'$ requires a better precision of
  measuring $m_c$ in each event. However, we also note that for a
  given dark matter model, the amount of charge stored in each neutron
  star, and thus $\tilde\al'$, can be estimated to a certain
  extent, which can in turn be translated to a bound on the
  properties of aDM in NS. 
In the case of LIGO, aDM fraction can be constrained starting from the
heavy end in the $m_\chi \sim \mathrm{TeV}$ range. With very heavy aDM, the
fraction can be constrained down to sub-percentage level. 
%
  This is to be compared with
  5-10\% DM mass allowed in NS in models such as
  \cite{Foot:2004pa,Sandin:2008db,Fan:2013yva}, according to
  \cite{Ellis:2017jgp}, which assumes a different DM model.
We also note,  the bound on $\tilde\al'$ for a specific dark
  matter model can be improved with increasing the statistics. 


The second window comes from the GW wavelength, to which LIGO-Virgo is sensitive in the band $10^3\;\text{km}<\lambda_{GW}<10^4$ km. Observation of hidden sector in this window relies on a nonzero charge-mass difference $\ga$ for the two NS. When this is the case, the dark dipole radiation is stimulated when $\lam_{GW}\lesssim m_V^{-1}$, and at this point the waveform will develop a dipole component which is vastly different from all the other contributions. For a very light dark photon with $m_V^{-1}\gg 10^4$ km, such that $\ga\neq 0$ in the whole observational window, the dipole component exists throughout the LIGO-Virgo band. This is illustrated in Figure~\ref{fig:freq-time-diff-mV}.

We conclude by noting the following points. The existence of NS
binaries itself puts a constraint on the percentage of aDM each NS
carries, which is equivalent of the weakest bound in our analysis by setting
$E=0$. This sets the same bound that is required for the aDM to be
contained in the NS. In addition, in this Letter we make no assumption
about the structure of the dark sector or the origin of the dark boson
mass. In the scenario where the dark photon mass is generated by a dark Higgs mechanism rather than a Stueckelberg mass term, the
signal will likely be affected by the dark Higgs, which we leave for future work. 

{ Similarly, we make no assumption about the mechanism of DM capture. All we assumed is that the DM mass fraction in a NS is large enough to generate an effect, while the coupled system of NS and DM remains stable. In the minimal scenario with dark repulsion the only long range force besides gravity, the stability implies that $\tilde\al'$ is always smaller than DM mass fraction. Therefore, a sensitivity of $\tilde\al'$ down to $10^{-2}$ means that percent-level mass fraction of DM in a NS is being constrained. We do not address how the DM got captured, but note  that for instance, DM could have been captured by the adiabatically contracting gravitational potential well during the formation of the progenitor star.  The above stability constraint between $\tilde\al'$ and DM mass fraction can be easily circumvented by, e.g., the presence of another   force for DM  which is attractive and has range $\lesssim\order{10}$km. 

}
Furthermore, if   mixing between the Standard Model photon and the dark boson is assumed, the dark radiation may leave imprints in the radio frequency band. Finally, our analysis here can be readily applied to GW detectors with different frequency bands such as LISA, extending the reach of the GW spectroscopy for hidden sectors.



\acknowledgments
The authors would like to thank Prateek Agrawal, Robert Caldwell, David Pinner, and Lisa Randall for insightful
discussions.  The authors would also like to thank the referee for
  constructive comments and relevant
  reference. AN is funded in part  by the DOE under grant
DE-SC0011637, and by the Kenneth K. Young Chair. CS is supported in part by the International Postdoctoral Fellowship funded by China Postdoctoral Science Foundation. ZZX is supported in part by Center of Mathematical Sciences and Applications, Harvard University. CS is grateful for
the hospitality and partial support of the Department of Physics and
Astronomy at Dartmouth College where this work was done.


\end{document}